%% file: main.tex
\documentclass[conference]{IEEEtran}

\input{head}

\usepackage{paralist}

\input{include}
\begin{document}
\bstctlcite{IEEEexample:BSTcontrol}

\title{RAPTOR: Ravenous Throughput Computing}

\author{Andre Merzky$^{1}$, Matteo Turilli$^{1,2}$, Shantenu Jha$^{1}$$^{,2}$\\
   \small{\emph{$^{1}$ Rutgers, the State University of New Jersey, Piscataway, NJ 08854, USA}}\\
   \small{\emph{$^{2}$ Brookhaven National Laboratory, Upton, NY 11973, USA}} \\
}

\maketitle

\begin{abstract}

We describe the design, implementation and performance of the RADICAL-Pilot task
overlay (RAPTOR). RAPTOR enables the execution of heterogeneous tasks---i.e.,
functions and executables with arbitrary duration---on HPC platforms, providing
high throughput and high resource utilization. RAPTOR supports the high
throughput virtual screening requirements of DOE's National Virtual
Biotechnology Laboratory effort to find therapeutic solutions for COVID-19.
RAPTOR has been used on $>8000$ compute nodes to sustain 144M/hour docking hits,
and to screen $\sim$10$^{11}$ ligands. To the best of our knowledge, both the
throughput rate and aggregated number of executed tasks are a factor of two
greater than previously reported in literature. RAPTOR represents important
progress towards improvement of computational drug discovery, in terms of size
of libraries screened, and for the possibility of generating training data fast
enough to serve the last generation of docking surrogate models.
\end{abstract}

\section{Introduction}\label{sec:intro}

\input{introduction}

\section{High-Throughput Virtual Screening}\label{sec:docking}
\input{docking}

\section{Design and Implementation}\label{sec:design}
\input{design}

\section{Performance Characterization}\label{sec:exp}

\input{experiments}

\section{Related Work}\label{sec:related}
\input{related}

\section{Discussion and Conclusions}
\label{sec:conclusion}
\input{conclusions}



\vspace{0.1in}

\noindent {\bf Acknowledgements} {We acknowledge Arvind
  Ramanathan, Rick Stevens, Austin Clyde (Argonne National Laboratory), and
  other members of the National Virtual Biotechnology Laboratory (NVBL)
  Medical Therapeutics group. We acknowledge computing time via the COVID19
  HPC Consortium. We thank TACC for the opportunity for scaling runs during
  the TexaScale days. This work is supported by NSF-1931512
  (RADICAL-Cybertools), NSF-1835449 (SCALE-MS) and ECP CANDLE and ExaWorks.
  This research used resources at the Oak Ridge Leadership Computing Facility
  at the Oak Ridge National Laboratory, which is supported by the Office of
  Science of the U.S. Department of Energy under Contract No.
  DE-AC05-00OR22725. }


\footnotesize


\bibliographystyle{IEEEtran}
\bibliography{radical,raptor,turilli-references}

\end{document}

%% file: head.tex
\usepackage[export]{adjustbox}
\usepackage{booktabs}
\usepackage{float}
\usepackage{flushend}
\usepackage{graphicx}
\usepackage{grffile}
\usepackage{ifpdf}
\usepackage[utf8]{inputenc}
\usepackage{listings}
\usepackage{multirow}
\usepackage[caption=false,font=footnotesize]{subfig}
\usepackage{url}
\usepackage{tabularx}
\usepackage{textcomp}
\usepackage{xcolor}
\usepackage{xspace}
\usepackage{cite}
\usepackage{comment}
\usepackage{outlines}
\usepackage{tikz} 

\usepackage[pagewise, modulo]{lineno}

\definecolor{listinggray}{gray}{0.95}
\definecolor{darkgray}{gray}{0.7}
\definecolor{commentgreen}{rgb}{0, 0.4, 0}
\definecolor{darkblue}{rgb}{0, 0, 0.6}
\definecolor{purple}{rgb}{0.6, 0, 0.6}
\definecolor{middleblue}{rgb}{0, 0, 0.75}
\definecolor{darkred}{rgb}{0.4, 0, 0}
\definecolor{brown}{rgb}{0.5, 0.5, 0}
\definecolor{dkgreen}{rgb}{0,0.5,0}
\definecolor{orange}{rgb}{1,.5,0}
\definecolor{dandelion}{cmyk}{0,0.29,0.84,0}

\usepackage[normalem]{ulem}
\makeatletter
\def\cyanuwave{\bgroup \markoverwith{\lower3.5\p@\hbox{\sixly \textcolor{cyan}{\char58}}}\ULon}
\def\reduwave{\bgroup \markoverwith{\lower3.5\p@\hbox{\sixly \textcolor{red}{\char58}}}\ULon}
\def\blueuwave{\bgroup \markoverwith{\lower3.5\p@\hbox{\sixly \textcolor{blue}{\char58}}}\ULon}
\font\sixly=lasy6 
\makeatother

\def\BibTeX{{\rm B\kern-.05em{\sc i\kern-.025em b}\kern-.08em
    T\kern-.1667em\lower.7ex\hbox{E}\kern-.125emX}}

%% file: include.tex
\newif\ifdraft{}
\drafttrue{}

\ifdraft{}
  \newcommand{\amnote}[1]{ \textcolor{blue} { ***andrem: #1 }}
  \newcommand{\jhanote}[1]{ {\textcolor{red} { ***shantenu: #1 }}}
  \newcommand{\mtnote}[1]{ {\textcolor{orange} { ***matteo: #1 }}}
  \newcommand{\fixme}{{\textcolor{red}{FIXME \xspace}}}

\else
  \newcommand{\amnote}[1]{}
  \newcommand{\jhanote}[1]{}
  \newcommand{\mtnote}[1]{}
  \newcommand{\fixme}[1]{}
\fi

\newcommand{\I}[1]{\textit{#1}\xspace}
\newcommand{\T}[1]{\texttt{#1}\xspace}

\newcommand{\UP}{\vspace*{-1.0em}}
\newcommand{\up}{\vspace*{-0.5em}}

\lstdefinestyle{myListing}{
  frame=single,
  backgroundcolor=\color{listinggray},
  language=C,
  basicstyle=\ttfamily \footnotesize,
  breakautoindent=true,
  breaklines=true
  tabsize=2,
  captionpos=b,
  aboveskip=0em,
  belowskip=-2em,
}

\lstdefinestyle{myPythonListing}{
  frame=single,
  backgroundcolor=\color{listinggray},
  language=Python,
  basicstyle=\ttfamily \footnotesize,
  breakautoindent=true,
  breaklines=true
  tabsize=2,
  captionpos=b,
}

\ifpdf{}
  \DeclareGraphicsExtensions{.pdf, .jpg, .tif}
\else
  \DeclareGraphicsExtensions{.eps, .jpg, .ps}
\fi

\newcommand*\circled[1]{\tikz[baseline=(char.base)]{
    \node[shape=circle,draw,inner sep=1pt] (char) {#1};}}

%% file: introduction.tex
In response to COVID19, many researchers are using high-performance computing
(HPC) to support epidemiological studies and to design anti-viral therapeutics.
Significant effort has been invested in designing drug-discovery pipelines that
can screen many more ligands than traditional \textit{in-silico} drug-design
approaches.

Nearly all high throughput virtual screening (HTVS) pipelines involve docking,
i.e., the process of scoring a putative drug candidate (ligand) with a potential
protein target. Docking algorithms are significantly cheaper but less accurate
than full physics-based simulations to compute binding affinities between ligand
and protease. The relative inexpensive scoring allows many more ligands to be
investigated, which is necessary given the possible $10^{60}$ drug candidates.

One noteworthy COVID19 HTVS pipeline is the IMPECCABLE
campaign~\cite{lee2020scalable,saadi2020impeccable}---DoE's National Virtual
Biotechnology Laboratory (NVBL)~\cite{nvbl} effort to develop therapeutics for
COVID-19. We discuss design, implementation and performance of RADICAL-Pilot
task overlay (RAPTOR), which serves as the workhorse of the campaign's docking
effort.

RAPTOR is a general purpose, portable, coordinator/worker framework for the
execution of function and executable tasks on HPC platforms. RAPTOR is a
subsystem of RADICAL-Pilot (RP)~\cite{merzky2021design}, and relies on it to
acquire and manage resources, and to schedule and launch its coordinator and
workers on those resources. RAPTOR extends RP's capabilities, providing: (1)
steady utilization above 90\% of the available resources with task executing for
1 second or longer; (2) partitioning of CPU and GPU resources across an
arbitrary number of concurrently and/or sequentially executing batch jobs; and
(3) partitioning of tasks across multiple, independent coordinators and workers.

To get a sense of the scale and impact: RAPTOR has been used on up to 466,816
concurrent CPU cores to sustain $144\times10^6$/hour docking
hits~\cite{clyde2021high} and to screen approximately 10$^{11}$ ligands. To the
best of our knowledge, both throughput rate and aggregated number of executed
tasks are a factor of two greater than previously reported in
literature~\cite{gorgulla2020open}. RAPTOR was used to generate consensus and
ensemble scoring against protein targets, and to generate training data for
docking surrogate models~\cite{regression2020clyde,clyde2021protein} that are up
to 3--4 orders of magnitude faster than traditional docking
programs~\cite{saadi2020impeccable}. Finally, RAPTOR has been used for more than
2M node-hours on primarily TACC Frontera and ORNL Summit, to support the
identification of over 40 hits on COVID19 drug targets that are progressing to
advanced testing~\cite{clyde2021high,babuji2020targeting}.

The main contributions of this paper are a description of the design and
implementation of RAPTOR, and a performance evaluation of RAPTOR when used to
perform computational docking at scale, as part of an HTVS pipeline.
\S\ref{sec:docking} describes the HTVS use case and the state of the art
infrastructure for it. \S\ref{sec:design} discusses the design and
implementation of RAPTOR, showing how it extends RP to support HTVS.
\S\ref{sec:exp} discusses experimental insight into the performance of RAPTOR
for different workloads and configurations used to run HTVS.

%% file: docking.tex

\begin{figure}
    \includegraphics[width=0.4\textwidth]{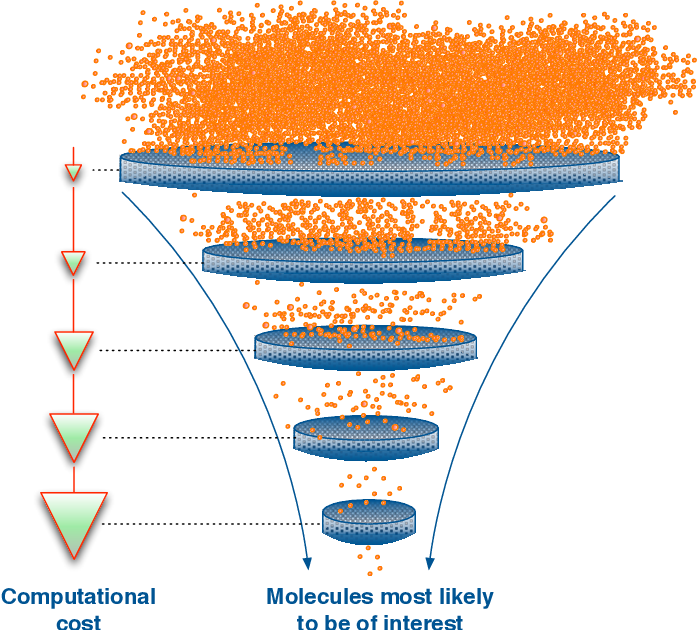}
    \up\up
    \caption{A schematic of high throughput virtual screening (HTVS): Downstream
    stages are progressively computationally more expensive, but are focussed on
    increasingly promising candidates \textit{Image source Ref.~\cite{pyzer2015high}}.}\label{fig:vs}
    \UP\up
\end{figure}

HTVS is used in a variety of disciplines, from materials
design~\cite{pyzer2015high} to drug discovery~\cite{zhang2008dovis}. HTVS
analyzes libraries of molecules, reducing them to a set of promising
leads for experimental evaluation. Typically, HTVS follows a computational
funnel (Fig.~\ref{fig:vs}) to focus computational effort on promising molecules.
Specifically, HTVS intelligently samples the large space of possible candidates
and narrows the number of candidates down to an experimentally tractable set.
HTVS is necessary for problems where exhaustive exploration is not an option, or
for time critical options where random search is not acceptable.

In drug-discovery, HTVS enables rapid, low-cost screening of significantly
larger and curated compound libraries, than feasible in experimental
studies~\cite{zhang2008dovis}. HTVS can now outperform equivalent experimental
high throughput screening, and has been shown to rapidly identify tightly
binding compounds. However, virtual libraries used in molecular discovery are
often still too large to exhaustively evaluate, warranting the use of algorithms
to help with exploration.

Computational docking often forms the first stage in an HTVS pipeline.
Docking fits trial drug-like compounds into (protein) binding sites in
three-dimensional models of the protein targets characterized by a score.
Docking is useful in early stages of molecular discovery to identify initial
hits to be prioritized for experimental validation. This is true for both
``regular'' drug discovery pipelines, as well as for customized and AI-based
ones.

\subsection{HTVS Docking Infrastructure: State-of-the-Art}

Several efforts created an open HTVS infrastructure, taking advantage of cloud
platforms~\cite{gorgulla2021multi,gorgulla2020open} or HPC
resources~\cite{acharya2020supercomputer} to support large-scale ligand docking
across various protein targets. Here we discuss four recent publications that
represent the spectrum of performance considerations and design properties.
Impressive
results notwithstanding, given the diverse computing platforms and docking
programs employed, as well as different measures of performance, it is difficult
to provide a head-to-head performance comparison.

\begin{inparaenum}

\item VirtualFlow~\cite{gorgulla2021multi} submits multiple ``jobs'' to
different ``clusters''. VirtualFlow exhibited linear scalability with a peak at
160,000 CPUs on Google Cloud. VirtualFlow 
can dock 1B compounds in approximately 2 weeks, using 10,000 CPU cores
simultaneously. In the context of COVID19, Ref.~\cite{gorgulla2021multi}
investigate 17 virus-related targets, 45 screens, and $\approx 50 \times 10^9$
docking instances. However, they do not report how effectively the resources are
utilized and focus on application performance measures (docks/time).

\item  Ref.~\cite{acharya2020supercomputer} outlines a supercomputer-driven
pipeline for \textit{in silico} drug discovery, using enhanced sampling
molecular dynamics (MD) and ensemble docking. Ensemble docking makes use of MD
results by docking compound databases into representative protein binding-site
conformations, thus taking into account the dynamic properties of the binding
sites. On Summit, that pipeline docked 1B compounds in 24 hours, using
Autodock-GPU~\cite{autodock-gpu}. In Fig. 6 of
Ref.~\cite{acharya2020supercomputer}, the authors outline the variation in
docking time, of both GPU-based and ``regular'' CPU-based docking programs,
observing fluctuations of 2x and 10x respectively.

\item Ref.~\cite{glaser2021high}, an ACM 2020 Gordon Bell Finalist, reported a
peak performance of 20,000 docks/second on Summit, i.e., an aggregated
performance of 50M docks/hour, for up to 20 poses per dock. This was primarily
achieved through adaptation of AutoDock-GPU\@: GPU offloading feature
calculations in re-scoring and database queries. A further 10x performance
improvement was achieved by using parallel database methods. So far, 70M
docks/hour is the highest reported throughput in literature.

\item The COVID-Moonshot project~\cite{achdout2020covid} used the folding@home
approach in conjunction with high-throughput experimental screening, MD
simulations and ML to identify covalent and non-covalent inhibitors against main
protease (MPro) which demonstrated viral inhibition. Although folding@home is an
``exascale'' platform, the peak or sustained throughput is not precisely
reported, nor is it easy to discern from available data. Based upon published
science results, our best-effort estimate is that the COVID-Moonshot project
screened $10\times 10^{9}$ ligands over a period of several months, using steady
state resources, thus about $100\times 10^{6}$ docks/day over approximately 1000
CPUs.

\end{inparaenum}

\subsection{RAPTOR for HTVS Docking}

RAPTOR is used for the docking phase of DoE NVBL's IMPECCABLE
campaign~\cite{saadi2020impeccable,lee2020scalable,nvbl,clyde2021high}, which
integrates algorithmic and methodological innovations with advanced
infrastructure to dock a large number of ligands with protein targets. A protein
target represents a well-defined binding site, expressed as PDB file. For each
target, we iterate through ligands from certain molecule libraries and compute a
docking score for that ligand-protein pair. To increase the reliability of
results, we used OpenEye and Autodock-GPU for the same ligand set and targets,
which also allowed us to leverage HPC resource heterogeneity. We executed
OpenEye on Frontera's x86 architectures and Autodock-GPU on Summit's GPUs. A
docking call is executed either as a task function of the OpenEye Python
library, or as an executable task launching AutoDock-GPU\@. In both cases,
RAPTOR is used for orchestration.

Docking was used for consensus and ensemble scoring of large
libraries~\cite{babuji2020targeting}, and to generate training data for docking
surrogate models~\cite{regression2020clyde}. Several libraries, the largest of
which (\texttt{mcule\--ultimate\--200204\--VJL}) has 126M drug candidates, were
used to dock against more than 100 targets. Other libraries include
\texttt{Orderable\--zinc\--db\--enaHLL} with 6.6M candidates, details of which
can be found in Ref.~\cite{clyde2021high}. Based upon library sizes, we estimate
RAPTOR has been used to screen close to $100 \times 10^{9}$ molecules against
over a dozen drug targets in SARS-CoV-2.

RAPTOR is a general-purpose, high-throughput task execution system that is not
limited to a specific docking program or computing platform. This is in contrast
to Refs.~\cite{gorgulla2021multi,achdout2020covid} and
Refs.~\cite{glaser2021high,vermaas2020supercomputing,acharya2020supercomputer},
respectively. Since the docking programs are different, a direct comparison of
scientific docking results are not meaningful, however RAPTOR reached a
throughput of 144M docks/h and performed a total of $100\times 10^9$ docking
tasks. These are both factor of two greater than previously reported in
literature. RAPTOR achieved this on Frontera, using 8300 nodes at peak and with
resource utilization above 90\%. Further, different from
VirtualFlow~\cite{gorgulla2020open,gorgulla2021multi}, RAPTOR can manage the
entire workload within a single large job, or spread across multiple jobs.

%% file: design.tex


The RADICAL-Pilot Task OveRlay (RAPTOR) is a coordinator/worker framework for
the execution of function and executable tasks on HPC platforms. It is designed
to enable high-rate task execution at scale, e.g., up to $144\times10^6tasks/h$
on TACC Frontera's 8300 nodes. RAPTOR is a component of RADICAL-Pilot
(RP)~\cite{merzky2021design} and relies on RP to acquire and manage resources,
and to schedule and launch its coordinators and workers on those resources.
RAPTOR is implemented in Python as is RP\@.

\begin{figure}
    \centering
    \includegraphics[width=\columnwidth,trim=1mm 0 4mm 1.5mm,clip]{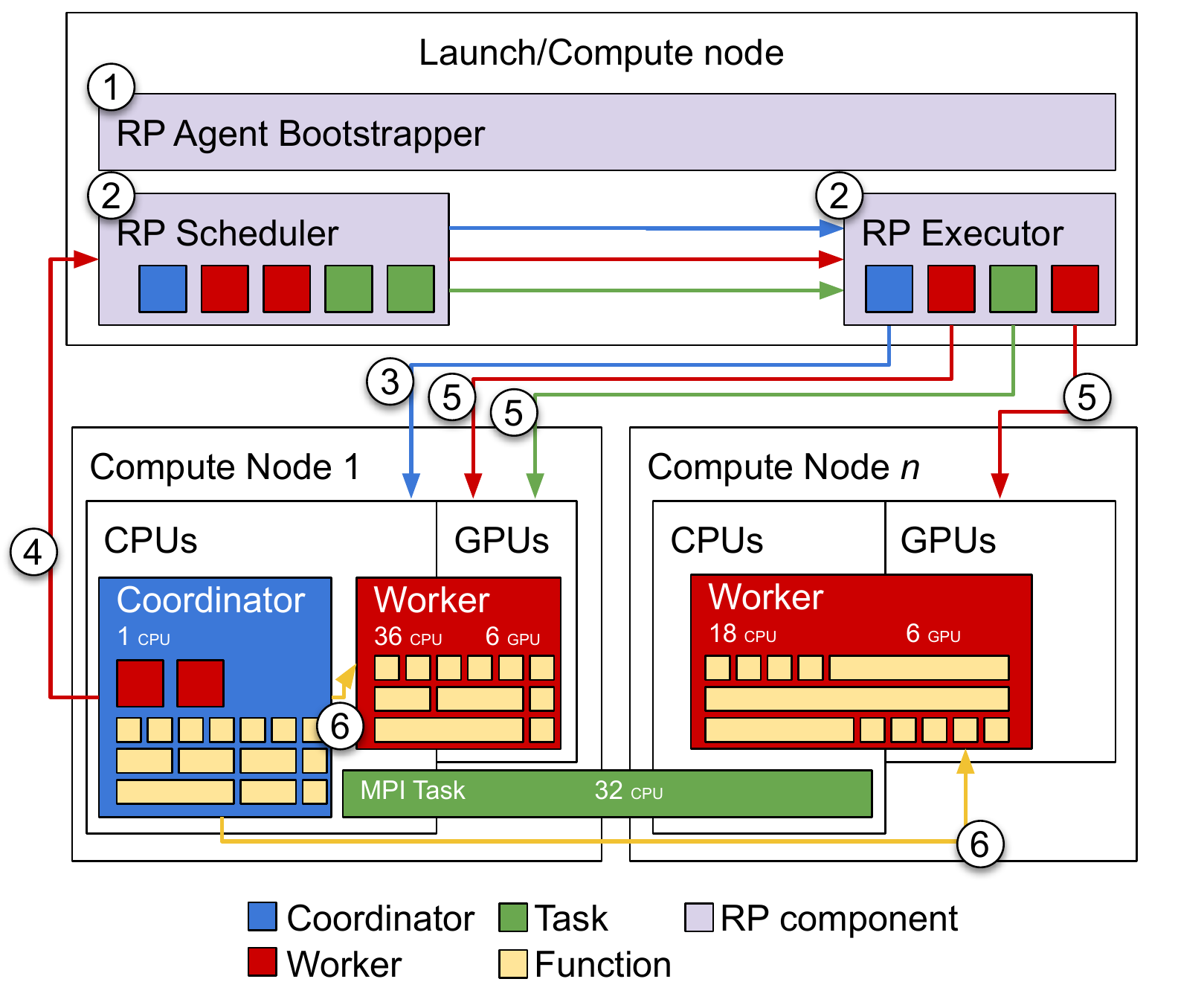}
    \UP
    \UP
    \caption{RAPTOR: architecture and execution model.}
    \label{fig:raptor}
    \UP
    \UP
\end{figure}

Fig.~\ref{fig:raptor} shows RP's architecture and execution model. RP exposes an
API to describe pilots and tasks~\cite{turilli2018comprehensive}, and uses four
modules to manage them: PilotManager, TaskManager, Agent and
DB~\cite{merzky2021design}. Once described, pilots and tasks are passed to RP's
runtime system \circled{1}. The PilotManager submits pilots on one or more
resources via the SAGA API \circled{2}. The SAGA API implements an adapter for
each supported resource type, exposing uniform methods for job and data
management~\cite{merzky2015saga}. Once a pilot becomes active on a resource, it
bootstraps the Agent module \circled{3}.

\begin{figure}
    \centering
    \includegraphics[width=\columnwidth,trim=0 0 0 0,clip]{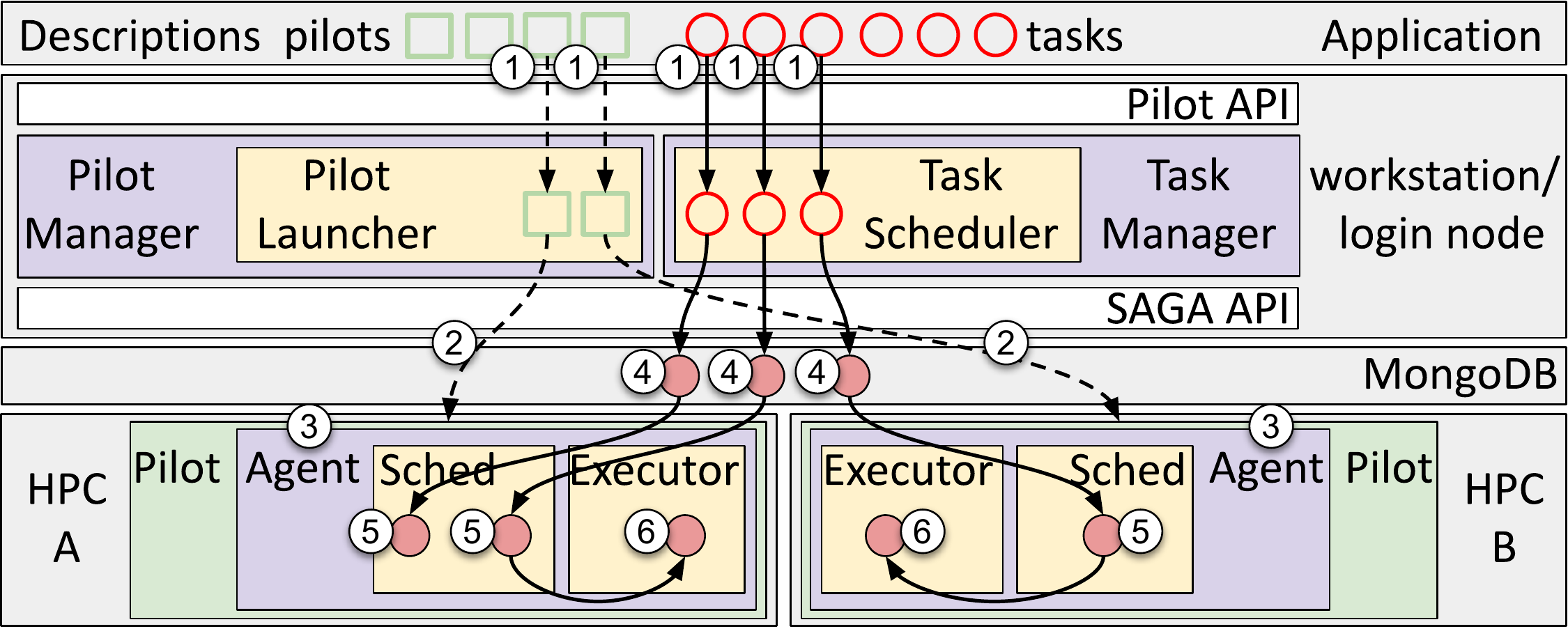}
    \caption{RADIAL-Pilot (RP) architecture and execution model. RP is a
    distributed system that execute its components between the user's
    workstation or the HPC platform's login node, and the HPC platform's compute
    nodes. RAPTOR executes on the target HPC platform's compute nodes (see
    Fig.~\ref{fig:raptor}).}\label{fig:exec-model}
  \end{figure}

The TaskManager schedules each task to an Agent \circled{4} via a queue on a
MongoDB instance. This instance is used as the RP DB module to communicate task
descriptions between the TaskManager(s) and the Agent(s). Each Agent pulls tasks
from the DB module \circled{5} and schedules \circled{6} each task on an
Executor upon resource availability (e.g., number of cores or GPUs). The
Executor sets up the task's execution environment and then spawns the task for
execution.

RP's tasks are fully-decoupled, i.e., they have no data dependences. Task
dependences can be resolved before submission to RP via workflow engines, e.g.,
EnTK~\cite{balasubramanian2018harnessing} and~\cite{babuji2019parsl}. Each task
is assumed to be self-contained, executed by RP as a black-box that returns
success or failure codes. RP has no control or knowledge about the code each
task executes, enabling the separation of concern among resource management,
execution management and task executables. At application level, RP implements a
`batch-like' programming model to describe groups of tasks (i.e., workloads) and
submit them for execution. Concurrency is implicit as users do not control it:
RP executes tasks with the maximum concurrency allowed by the available
resources.

RP is designed to schedule and launch executable tasks and not function tasks,
i.e., tasks coded as functions in a specific programming language. Executable
tasks are comprised of a self-contained program that can execute on the compute
nodes' operating system. RP handles such tasks as an object containing its
requirements, e.g., number of processes, type of process communication, type and
number of CPU cores and/or GPUs, and so on. RP's tasks are relatively `heavy'
and require a certain time to be scheduled and launched. That limits RP's
throughput and, ultimately, the efficiency at which RP can use resources with
tasks shorter than 1 minute~\cite{merzky2018using,merzky2021design}.

Scheduling in RP is global: all the tasks that are submitted to RP's Agent are
managed by a single scheduler. While the scheduling algorithm is tweaked to
reach peaks of 350 tasks/s, its performance degrades for short running tasks on
large resources (less than $\sim60s$ for $\sim1000$ nodes, $\sim120s$ for
$\sim2000$ nodes, etc.).

RAPTOR extends RP to support the execution of tasks like those required by the
COVID19 campaign described in~\S\ref{sec:docking}. RAPTOR can: (1) execute both
function and executable tasks; (2) achieve high throughput with arbitrary short
running tasks; (3) arbitrarily partition resources and tasks; (4) use a
multilevel scheduling in which workloads are partitioned and then subsets of
tasks are scheduled to subset of resources; and (5) partition tasks across
multiple, independent executors.

Fig.~\ref{fig:raptor} shows how RAPTOR integrates within RP to enable the setup
of the coordinator/worker infrastructure and how it launches and executes tasks
on it. Due to RP's task model, scheduling and launching RAPTOR's coordinators
and workers do not require additional capabilities: once bootstrapped,
\circled{1} and \circled{2}, RP manages coordinators and workers as any other
task \circled{3}. Once running, a coordinator schedules one or more workers on
RP's Scheduler \circled{4}. Each worker is then launched on a compute node by RP
Executor \circled{5}. Finally, the coordinator schedules function calls on the
available workers for execution \circled{6}, load-balancing across workers as to
obtain maximal resource utilization.

As a subsystem of RP, RAPTOR integrates with overall RP's capabilities. RP's
TaskManager schedules and launches RAPTOR's coordinators and workers but also
other executable tasks. In turn, RAPTOR can execute both function and executable
tasks on the resources acquired by RP's PilotManager and allocated to RAPTOR's
workers, independent of whether RP executes tasks on other resources. This is
illustrated in Fig.~\ref{fig:raptor} by showing RP executing a 32 CPU MPI task
(green) on two compute nodes, while RAPTOR executes Python function calls of
diverse sizes (yellow) on its workers (red).

RAPTOR's coordinators and workers manage the execution of tasks via several
queues, depending on configuration parameters and performance requirements. A
coordinator pushes tasks to a queue and N workers concurrently pull that queue
for tasks to execute. The number of coordinators, queues and workers can be
tuned so that the rate of (de)queuing does not exceed the capabilities of the
queue implementation and of the used network. That keeps resources busy and
avoid worker starvation. Both coordinator and worker are implemented in Python,
using ZeroMQ to implement their queues.

RAPTOR's performance mainly depends on set up and management times of
coordinators and workers. Five design choices improve performance: (1) launch
coordinators and workers via MPI so to reduce latency, given the available
technologies; (2) use a dedicated communication channel between each
coordinator, its workers, and RP's scheduler; (3) partition resources across a
user-defined number of coordinators and workers; (4) limit each worker to use at
most one compute node; and (5) submit function tasks in bulk from a coordinator
to its workers.

RAPTOR's design enables the implementation of workload control at coordinator
level. RAPTOR coordinator's API consists of the \texttt{rp.raptor.coordinator}
class and four main methods: \texttt{submit}, \texttt{start}, \texttt{join} and
\texttt{stop}. Users inherit the class and initialize their coordinator with
parameters like the workers' description (\texttt{dscr}), and the number of
workers (\texttt{n\_worker}), CPUs (\texttt{cpn}) and/or GPUs (\texttt{gpn}).
Users then specify the workers' payload---either a Python function or an
arbitrary executable---and, in case, callbacks to receive updates about the
workers' status. Class methods are then used to submit the payload to the
workers (e.g., \texttt{coordinator.submit(descr\=descr, \ldots)}).

RAPTOR limits workers to a single node, making impossible to execute multi-node
MPI Python function tasks. This is acceptable considering that RP would still be
able to execute multi-node MPI executable tasks alongside the
coordinators/workers tasks. As discussed in the next section, the maximum number
of coordinators and workers that RP can manage depends on the HPC platform's MPI
performance. RAPTOR's throughput depends on the workload executed and on system
capabilities such as shared file system performance, availability of caches,
etc. RAPTOR enables resource-specific optimizations (e.g., using nodes' SSD
storage) that are not necessarily portable due to hardware and system
constraints. Currently, RAPTOR provides workers for Python functions and code
snippets, and for arbitrary non-MPI executables. Prototypes of workers to
execute multi-node MPI functions and executables are being tested.

%% file: experiments.tex
We characterize RAPTOR's performance and overheads, showing how it supports
the execution of the workload described in~\S\ref{sec:docking} at scale.  We
perform experiments with: production runs for NVBL-Medical Therapeutics
campaigns on Frontera; runs for largest achievable size on both Frontera and
Summit; and runs with both function and executable tasks. Tab.~\ref{tab:exp}
summarizes the parameterization and results of each experiment. We measure:
docking time in seconds (s); docking rate in docks/h; resource utilization;
and RAPTOR set-up time in seconds.

Resource utilization measures the percentage of available CPU and/or GPUs used
for docking operations. Resources become available as soon as the HPC platform's
batch system schedules the job(s) submitted by RP\@. As such, resource
utilization is a measure of how efficiently RP and RAPTOR use the available
resources to execute the given workload. Extending the results presented in
Ref.~\cite{lee2020scalable}, tab.~\ref{tab:exp} provides two values for resource
utilization: \T{avg} for the average utilization over the pilot runtime, and
\T{steady} for the steady-state utilization. For the latter, we remove the
contributions of startup and cool-down. We define startup as the time where the
concurrency of tasks rises, and cool-down where the concurrency decreases.

\begin{table*}
  \caption{Experiments. RAPTOR uses one pilot for each protein, computing
  the docking score of a variable number of ligands to that protein.
  OpenEye and AutoDock-GPU implement different docking algorithms and docking
  scores, resulting in different task times and rates. Resource utilization is
  often impeded by the long tail task time distributions which cause an
  expensive cooldown period. However, the steady state resource utilization is
  $>=$90\% for all experiments.}\label{tab:exp}
  \centering
  \begin{tabular}{ c  
                   l  
                   l  
                   r  
                   r  
                   r  
                   r  
                   r  
                   r  
                   r  
                   r  
                   r  
                   r} 
  \toprule
  \multirow{2}{*}{\textbf{ID}}               &
  \multirow{2}{*}{\textbf{Platform}}         &
  \multirow{2}{*}{\textbf{Application}}      &
  \multirow{2}{*}{\textbf{Nodes}}            &
  \multirow{2}{*}{\textbf{Pilots}}           &
  \textbf{Tasks}                             &
  \textbf{Startup}                           &
  \textbf{1st Task}                          &
  \textbf{Utilization}                       &
  \multicolumn{2}{c}{\textbf{Task Time [$sec$]}}  &
  \multicolumn{2}{c}{\textbf{Rate [$\times10^6/h$]}}  \\
  \cline{9-12}
                          &   
                          &   
                          &   
                          &   
                          &   
  \textbf{[$\times10^6$]} &   
  \textbf{[$sec$]}        &   
  \textbf{[$sec$]}        &   
  avg / steady            &   
  \textbf{max}            &   
  \textbf{mean}           &   
  \textbf{max}            &   
  \textbf{mean} \\            
  \midrule
  1             &  
  Frontera      &  
  OpenEye       &  
  128           &  
  31            &  
  205           &  
  129           &  
  125           &  
  90\% / 93\%   &  
  3582.6        &  
  28.8          &  
  17.4          &  
  5.0           \\ 
  2             &  
  Frontera      &  
  OpenEye       &  
  7600          &  
  1             &  
  126           &  
  81            &  
  140           &  
  90\% / 98\%   &  
  14958.8       &  
  10.1          &  
  144.0         &  
  126.0         \\ 
  3             &  
  Frontera      &  
  OpenEye       &  
  8336          &  
  1             &  
  13            &  
  451           &  
  142           &  
  63\% / 98\%   &  
  219.0         &  
  25.3          &  
  91.8          &  
  11.0          \\ 
  4             &  
  Summit        &  
  AutoDock      &  
  1000          &  
  1             &  
  57            &  
  107           &  
  220           &  
  95\% / 95\%   &  
  263.9         &  
  36.2          &  
  11.3          &  
  11.1          \\ 
  \bottomrule
  \end{tabular}
  \UP\UP
\end{table*}

We assign one pilot for each protein to which a set of ligands will be docked.
Within each pilot, each coordinator will manage the workers defined via its
interface (see~\S\ref{sec:design}). Each coordinator iterates at different
strides through the ligands database, using pre-computed data offsets for faster
access, and generating the docking requests to be distributed to the workers.
Each worker runs on one node, executing docking requests across the CPU cores or
the GPUs of that node.

\subsection{Experiment 1}\label{ssec:exp1}

Experiment 1 performed the docking of $6.6 \times 10^6$ ligands---from the
\texttt{Orderable\--zinc\--db\--enaHLL} database---on 31 proteins. We used RP to
acquire resources via 31 independent pilots, i.e., 31 jobs submitted to
Frontera's batch system. We used 1 pilot for each protein and, for each pilot,
we used RP to initialize RAPTOR's coordinators and workers. Each coordinator
then managed the execution of the docking function tasks on its workers. Due to
the different queue waiting times, at most 13 pilots executed concurrently in
experiment 1, with a peak throughput of $\sim17.4\times10^6$ docks/h.

The number of pilots used depends on the policies that govern: (1) the number of
jobs that can be concurrently submitted to the batch job system; (2) the amount
of resources a batch job can request; and (3) the maximal walltime allowed for
each job. We used Frontera's normal queue for experiment 1 that allowed us up to
100 concurrent jobs, each with a maximum of 1280 nodes and 48h of walltime. For
experiments 2 and 3, we used instead a single pilot as we had access to a
special queue that spanned all the machine for 24 and 3 hours respectively.
Importantly, RP and RAPTOR required no further coding to support those diverse
running modalities but only the setting of some of their configuration
parameters.

Earlier runs with a setup similar to experiments 1 encountered performance
bottlenecks related to Frontera's shared file system which stalled the
simulation progress. To keep an acceptable load on Frontera's shared filesystem,
only 34 of the 56 cores available were used on each node in experiment 1. Also
in this case, no coding was required but just changing a configuration
parameter.

\begin{figure}
  \centering
  \subfloat[][]{
      \includegraphics[width=0.44\textwidth]{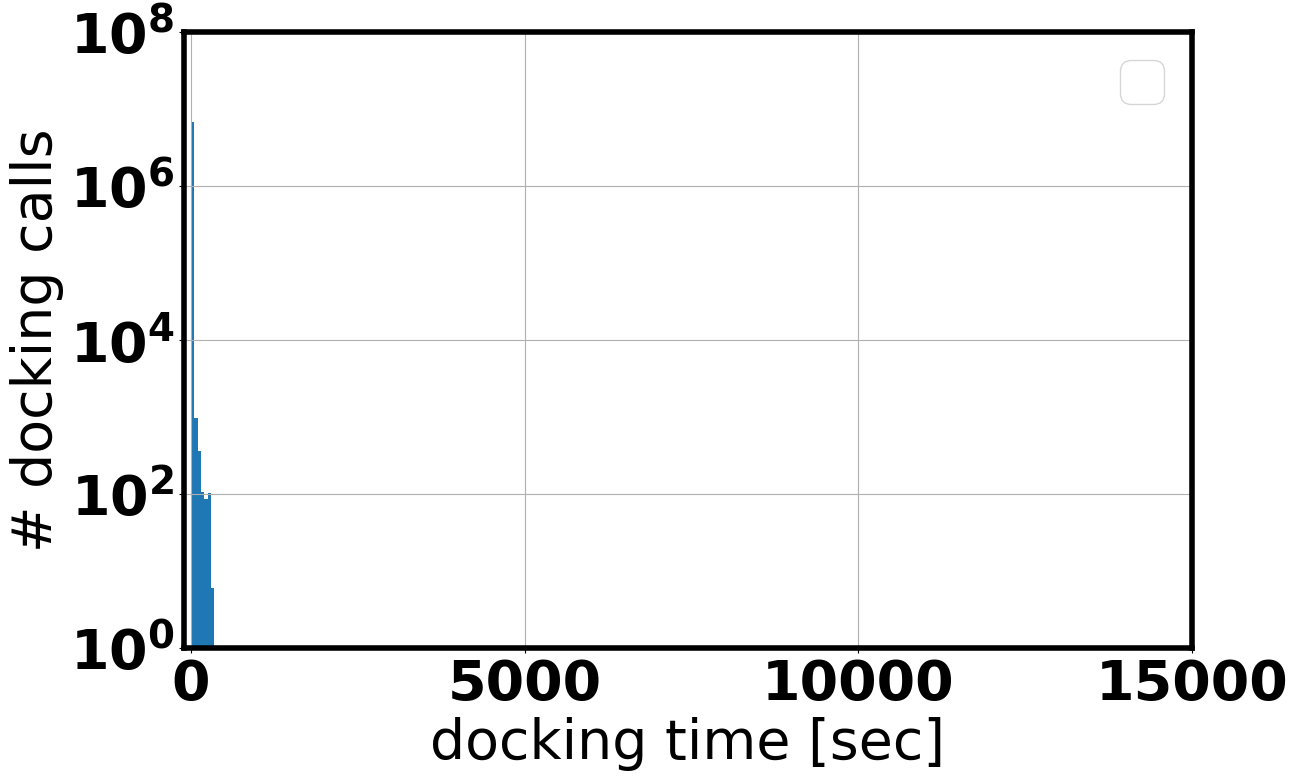}
      \label{sfig:exp1_durations_short}}
  \\
  \UP
  \subfloat[][]{
      \includegraphics[width=0.44\textwidth]{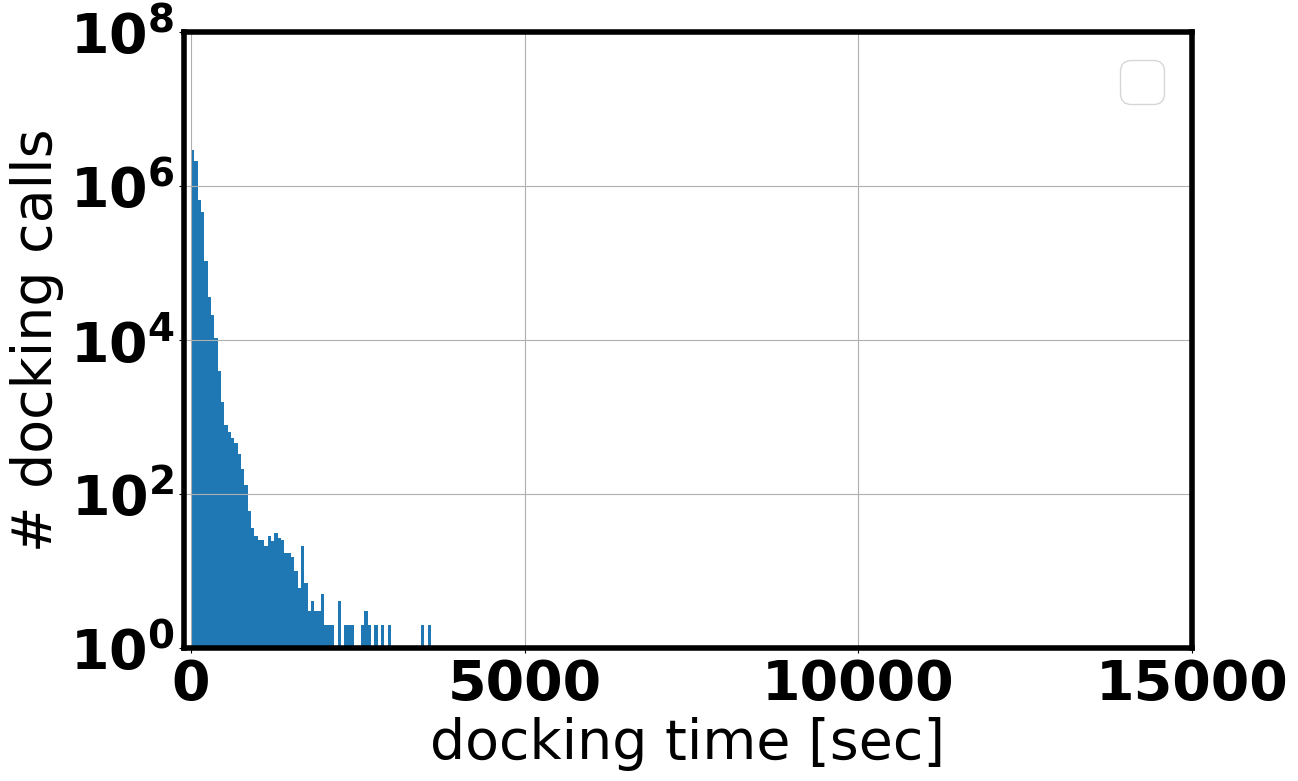}
      \label{sfig:exp1_durations_long}}
  \caption{Experiment 1: Distribution of docking times with the
      (a) shortest and (b) longest average time out of the 31 proteins analyzed.
      The distributions of the docking times for all 31 proteins have a long
      tail.}\label{fig:exp1_durations}
  \UP
  \UP
\end{figure}

Figs.~\ref{sfig:exp1_durations_short} and~\ref{sfig:exp1_durations_long} show
the distribution of docking times for proteins with the shortest and longest
average docking time, using the \texttt{Orderable\--zinc\--db\--enaHLL} ligand
database. All proteins are characterized by long-tailed docking time
distributions. Across the 31 proteins, the max/mean docking times are
3582.6/28.8s (Tab.~\ref{tab:exp}).

The large number of resulting docking requests ($31 \times 6.6 \times 10^6 = 205
\times 10^6$) poses a challenge to scalability due to the communication and
coordination overheads. The long tail distributions necessitate load balancing
across available workers to maximize resource utilization and minimize overall
execution time. Thus, docking requests cannot be assigned statically to workers,
but need to be dispatched dynamically.

Consistent with RAPTOR's design, we addressed load balancing by: (i)
communicating tasks in bulk so as to limit the communication frequency and
therefore overhead; (ii) using multiple coordinator processes to limit the
number of workers served by each coordinator, avoiding bottlenecks; (iii) using
multiple concurrent pilots to partition the docking computations.

\begin{figure}
  \centering
  \UP
  \subfloat[][]{
    \includegraphics[width=0.44\textwidth]{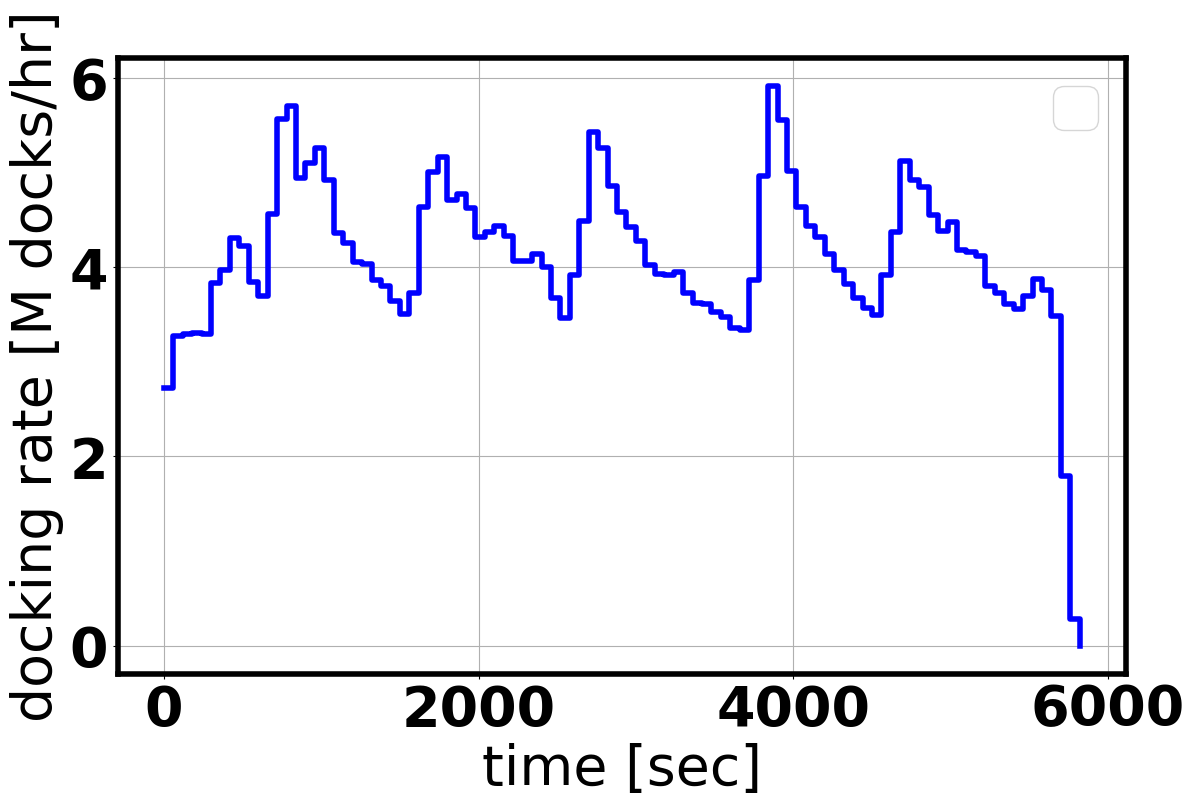}
    \label{sfig:exp1_rate_short}}
  \\
  \UP
  \subfloat[][]{
    \includegraphics[width=0.44\textwidth]{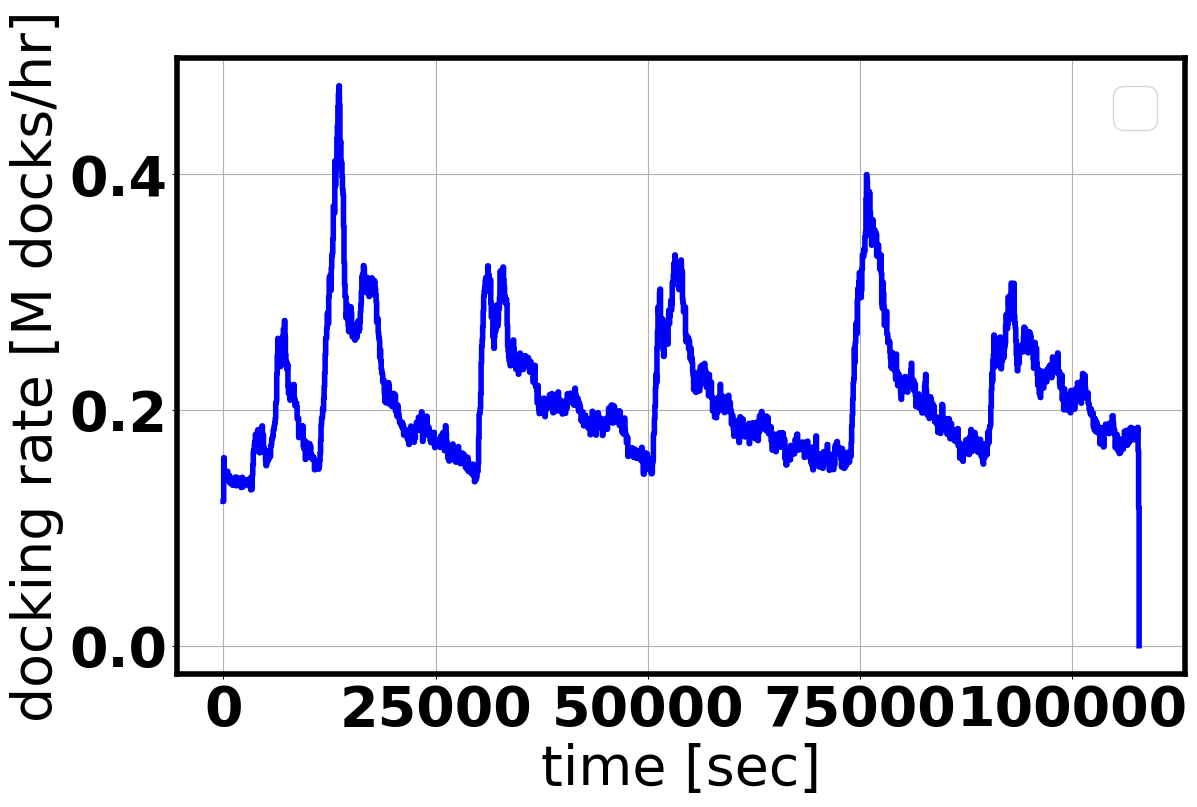}
    \label{sfig:exp1_rate_long}}
  \caption{Experiment 1: Docking rates for the protein with (a) shortest
           and (b) longest average docking time. The docking
           rates given in Tab.~\ref{tab:exp} are aggregated over concurrent
           pilots and thus larger than shown here for individual
           pilots.}\label{fig:exp1_rate}
  \UP
  \UP
\end{figure}

Figs.~\ref{sfig:exp1_rate_short} and~\ref{sfig:exp1_rate_long} show the docking
rates for the pilots depicted in Figs.~\ref{sfig:exp1_durations_short}
and~\ref{sfig:exp1_durations_long}, respectively. As with docking time
distributions, the docking rate is similar across proteins. It seems likely that
rate fluctuations depend on the interplay of machine performance, pilot size,
and specific properties of the ligands being docked, and the protein. We
measured a mean docking rate of $5\times10^6$ docks/h, with a max rate
$17.4\times10^6$ docks/h when 13 pilots where executing concurrently, using
about 20\% of Frontera's resources (Tab.~\ref{tab:exp}).

Note that each pilot needs some time to launch coordinators and workers, and to
begin distributing data and docking requests. Further, each pilot also needs
some time to terminate and collect trailing results. That behavior is visible in
all experiment plots. The respective overheads depend on the pilot size, and we
will discuss them in more details for a larger setup in experiment 3. On
Frontera, the time between when the pilot started and the first worker started
to execute the first task was $\sim$120s on average (Tab.~\ref{tab:exp}). That
includes $\sim$55s taken by the workers' coordinator to setup the execution and
the time needed to prepare some of the input parameters of the docking
functions. As the total execution time was between 1h:40m and 27h:46m
(Fig.~\ref{fig:exp1_rate}), setup time overhead was not relevant.
Tab.~\ref{tab:exp} also provides resource utilization (as defined above) for the
steady state between startup and cooldown, and as average for the full pilot
lifetime.

\subsection{Experiment 2}\label{ssec:exp2}

Experiment 2 characterizes the scalability of RAPTOR with a single pilot,
spanning all available nodes on Frontera (about 1000 nodes were reserved for
system work at the time of this run). To minimize the overhead caused by
repeated loading of the receptor data into memory, the data were loaded once per
node and then reused for all docking runs assigned to that specific node. The
individual cores hosting the docking computations received cloned copies of the
receptor data so as to isolate the individual docking computations.

To reduce the overhead of loading compound data from disk, the storage offsets
in the dataset were precomputed at startup and staged to the compute nodes.
Intermediate data were stored on node-local SSDs, further reducing the load on
the shared file system. For the same reason, during startup we stored a static
Python virtual environment with the OpenEye docking modules on the local storage
of the nodes. Together, those improvements enabled the use of all the 56 cores
of each compute node and required minor programming at application level. RP and
RAPTOR enable that kind of performance tuning for every application written
against their APIs and every HPC platform with local storage on the compute
nodes.

\begin{figure}
  \up
  \centering
  \subfloat[][]{
    \includegraphics[width=0.44\textwidth]{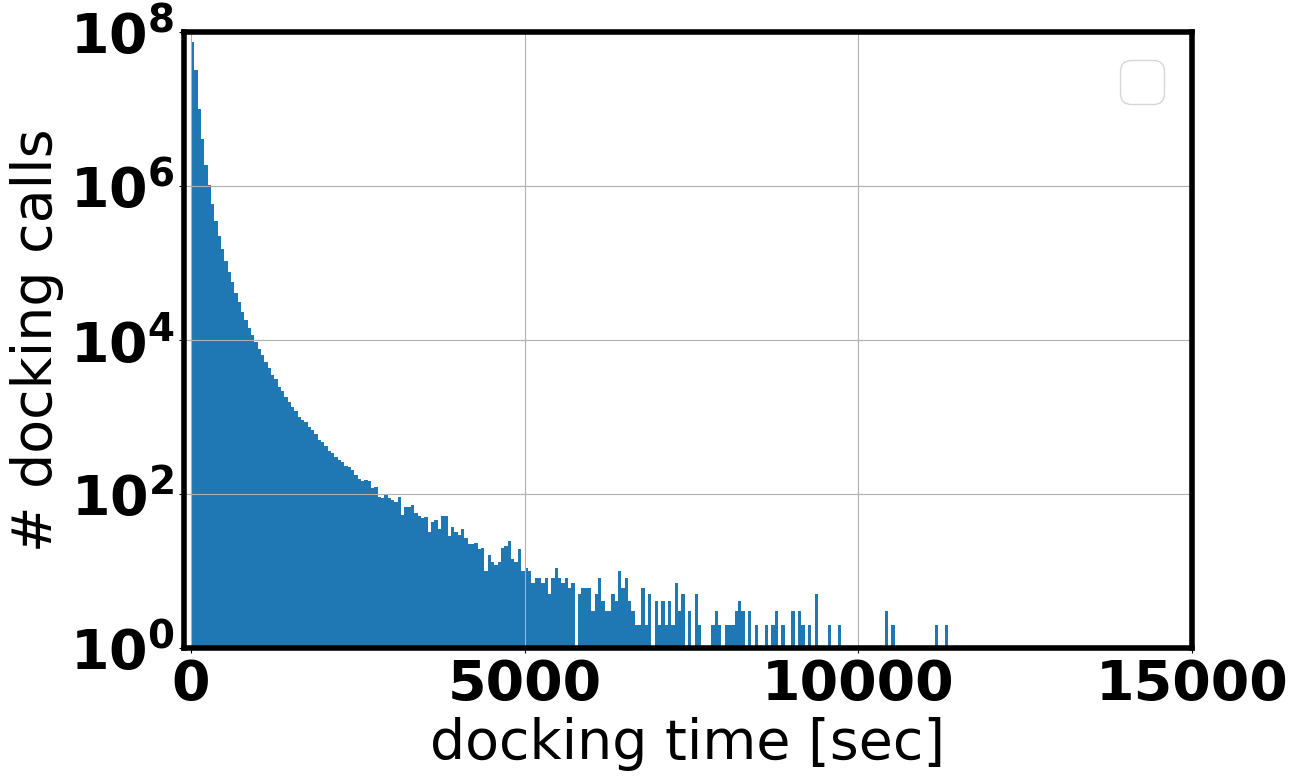}
    \label{sfig:exp2_time_ts2}}
  \\
  \UP
  \subfloat[][]{
    \includegraphics[width=0.44\textwidth]{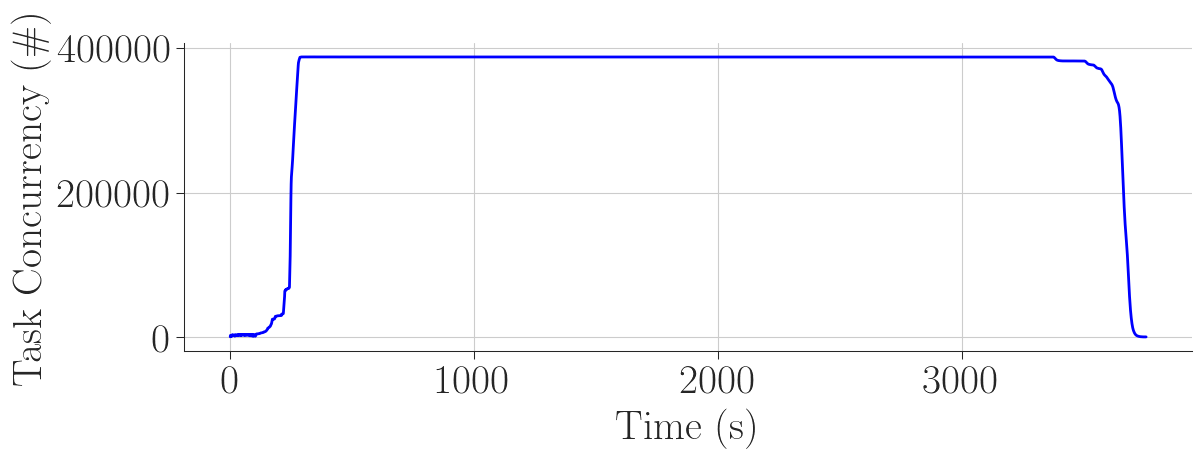}
    \label{sfig:exp2_conc}}
  \\
  \UP
  \subfloat[][]{
    \includegraphics[width=0.44\textwidth]{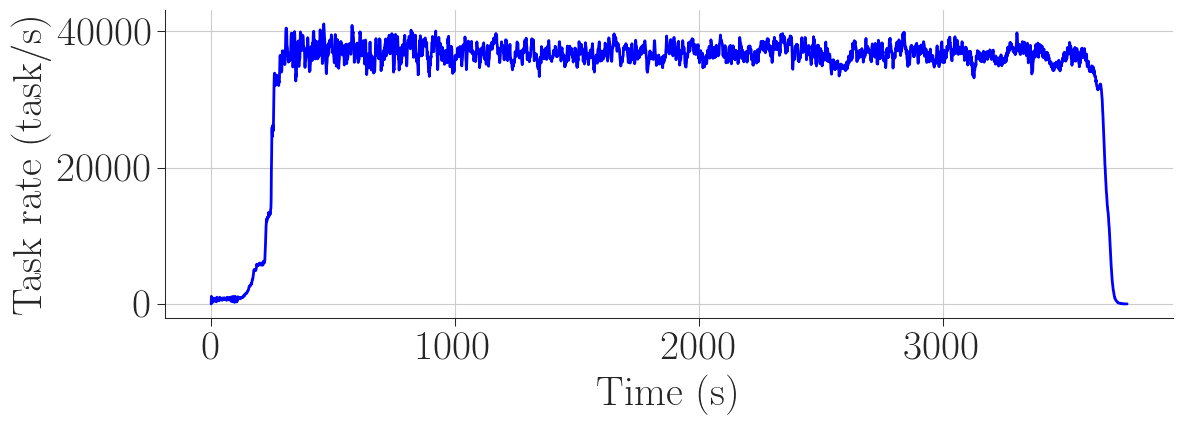}
    \label{sfig:exp2_rate_ts2}}

  \caption{Experiment 2: (a) docking time distribution; (b) docking
    concurrency; and (c) docking rate for a single protein and $126\times10^6$
    ligands. Executed with 158 coordinators, each using $\sim$50 nodes/2800
    cores on Frontera.}
  \label{fig:exp2_ts2}
  \UP
  \UP
\end{figure}

Fig.~\ref{sfig:exp2_time_ts2} shows the distribution of docking times of
approximately $126\times10^6$ ligands from the
\texttt{mcule\--ultimate\--200204\--VJL} library to a single protein, using
OpenEye on Frontera. Note that the distribution is highly dependent on the
protein being used: for the specific protein used in this run, we measured a max
of 14985.8 seconds and a mean of 61.5 seconds (Tab.~\ref{tab:exp}). The set of
proteins available to us varied in mean docking time from $\sim$3 to $\sim$70
seconds.

Fig.~\ref{sfig:exp2_rate_ts2} shows the docking rate for a single pilot with
7650 compute nodes. Compared to experiment 1, the rate does not fluctuate over
time and is consistently near $\sim40\times10^3$ docks/s ($\sim144\times10^6$
docks/h---see Tab.~\ref{tab:exp}). Note that the long tail distribution of
runtimes results in a long tail of docking calls and causes the ``cooldown''
phase. That phase and the startup time ultimately lower utilization from 98.3\%
in the steady-state (before cooldown starts) to a total average of 90.0\%. The
time taken to create the task was reduced to 35s from the 55s required in
experiment 1, mainly due to using the more efficient local storage. The time
taken to execute the first task from when resources become available marginally
increased to 140s compared to the 120s of experiment 1.

As discussed, the docking times depend on the proteins used. Thus, the docking
rate inversely depends on that protein choice and, ultimately, not on RAPTOR's
design and capabilities. The range of rates is very wide: for the proteins
available to us, we observed a mean docking rate between $\sim14\times10^6$ and
$\sim300\times10^6$, for runs of the same size.

\subsection{Experiment 3}\label{ssec:exp3}

One of RAPTOR's distinguishing capability is executing function and executable
tasks concurrently. In experiment 3, we launched executable tasks alongside the
OpenEye docking function tasks, thus emulating an heterogeneous workload. Each
executable task run the \texttt{stress} command.

As with experiment 2, we used only one pilot but with 8336 compute nodes, for a
total of 466,816 cores. On that pilot, we launched 8 coordinators and each
coordinator launched 1041 workers. As we use MPI to launch the workers, we
created a total of 8328 ranks, utilizing all the available compute nodes
(reserving 8 nodes for the coordinator's tasks). We tested larger numbers of
coordinators/workers but without measurable improvements. That implies that the
communication system is not a bottleneck in this setup. More experiments are
needed but, currently, we believe we reached the limit of the platform's
performance.

Executing experiments on whole machines of the size of Frontera requires special
agreements and support. We worked with Frontera staff at TACC to use the whole
machine for 3 hours after a maintenance period. Our runs uncovered issues with a
switch, triggered two faults in the shared file system and ultimately
overwhelmed the telemetry system of the machine. That reduced the time we had
for each run to 1200s and reduced the number of runs we were able to perform.

With short runs, the startup and cooldown periods have a relatively large impact
on the performance and utilization numbers presented here. Note that, in
production, users run RAPTOR for up to the maximal walltime allowed---usually
between 24 and 48 hours. RAPTOR startup time happens only once per run so, as
far as startup costs few minutes, it will be negligible from a RAPTOR's
efficiency point of view. On the contrary, filesystem slow downs and bottlenecks
recur across the whole run, greatly affecting overall resource utilization.
Thus, trading off slower startup time for better filesystem performance benefits
the overall run efficiency.

\begin{figure}
  \up
  \centering
    \subfloat[][]{
      \includegraphics[width=0.44\textwidth]{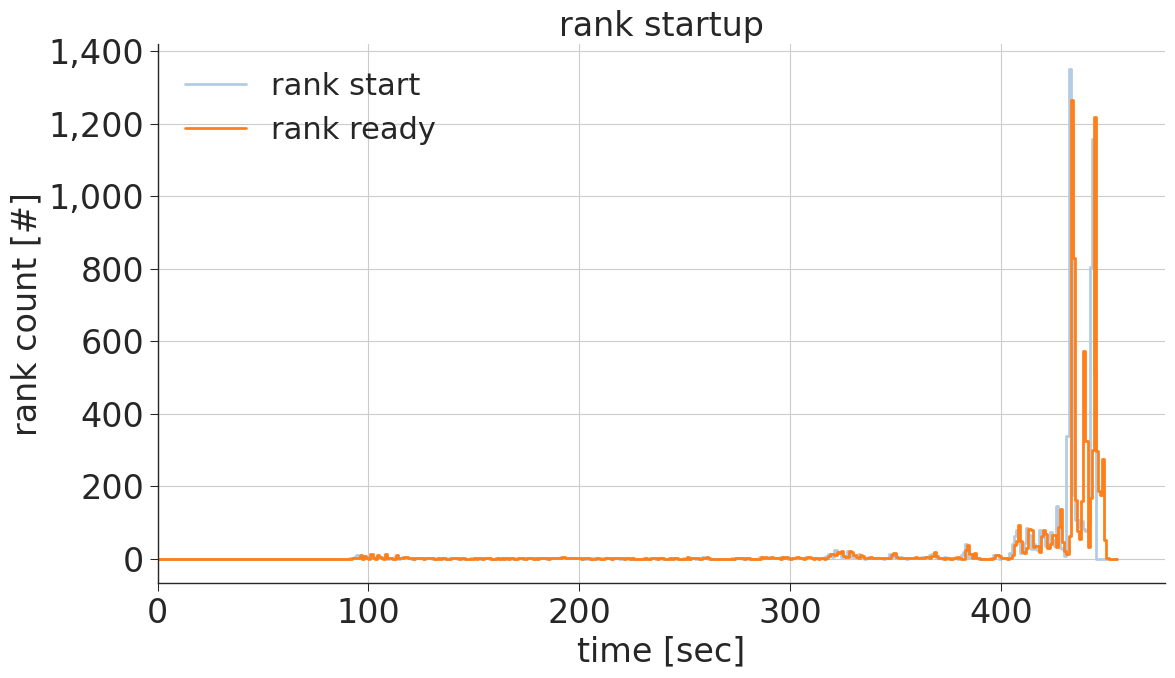}
      \label{sfig:exp3_startup}}
    \\
  \UP
    \subfloat[][]{
      \includegraphics[width=0.44\textwidth]{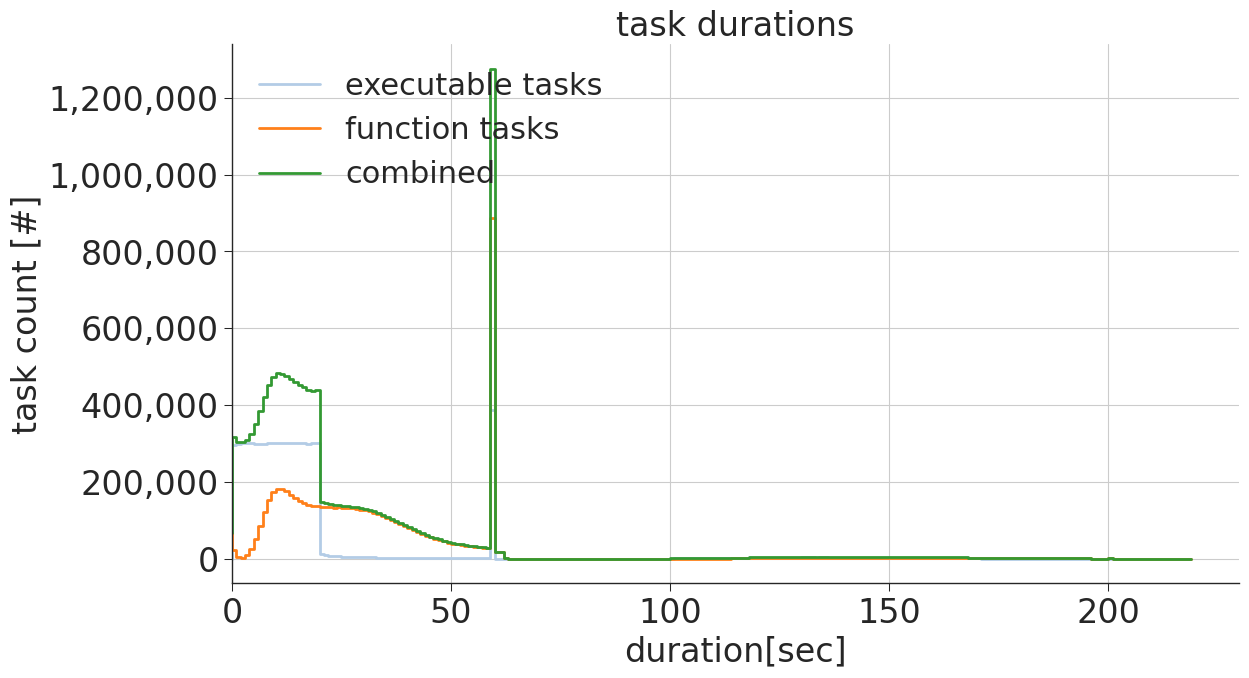}
      \label{sfig:exp3_runtimes}}
  \caption{Experiment 3: worker rank startup times.} 
  \label{fig:exp3_ranks}
  \UP
  \UP
\end{figure}

Despite the optimizations done to reduce the load on the shared file system in
experiments 2 and 3, most workers' task collection stalled for $\sim$150 seconds
after running as expected for $\sim$800 seconds. That stall led to longer task
runtimes visible in Fig.~\ref{sfig:exp3_runtimes}, where several tasks run
significantly longer beyond their nominal 60s cutoff time. The average
utilization listed in Tab.~\ref{tab:exp} was lowered significantly due to the
stalls. Our traces show no errors, delays or overloads and TACC telemetry
service did not provide conclusive data.

The startup time in experiment 3 is non-negligible: 451s. That amount of time
can be separated into 6 contributions: (1) pilot bootstrapping and (2) staging
to node storage overlap, contributing 78s; (3) coordinator startup that
contributes 1s; (4) input data pre-processing in the coordinators contributes
42s; (5) worker startup (all ranks) and (6) bootstrapping of communication
system overlap, contributing 330s.

Fig.~\ref{sfig:exp3_startup} shows a histogram of the startup times for all
ranks, and one for the communication channel setup which the worker ranks can
only initiate once the ranks are up. Interestingly, the first worker rank for
each coordinator took only $\sim$10 seconds to start, but the startup of the
remaining ranks was significantly slower, with the last worker to come alive
only after 330 seconds. These times depended on the performance of MPI on
Frontera and it calls for further optimization.

As soon as each coordinator and its workers become active, they started
executing bulks of 128 mixed function and executable tasks. The first worker
began executing tasks 142s after the job started; the last worker however
executed its first task at 368s, leading to a total ramp up time of 374s
observed in Fig.~\ref{sfig:exp3_startup}. Only then, RAPTOR could begin to
utilize the full system.

Fig.~\ref{sfig:exp3_runtimes} shows the task runtime distribution for this run
for 6,685,316 ligands from the \texttt{Orderable\--zinc\--db\--enaHLL} library
that are docked to the protein \texttt{3CLPro\--6LU7\--A\--1\--F} which is
particularly relevant to study the binding of drugs to the spike of SARS-CoV-2.
The figure shows durations between 3 and 60s and then a certain number of tasks
that have been terminated at 60s. This is the threshold used by the scientists
to determine when a ligand should be stopped to be computed, either because it
would not be relevant or because, more commonly, the simulation stalled.

Fig.~\ref{sfig:exp3_runtimes} also shows the distribution of the additional
6,685,316 \textbf{executable} tasks. We drew the tasks runtimes from a uniform
distribution between 0s and 20s. Note that both distributions show several tasks
running for longer than the 60s cutoff, up to a runtime of 360s. Those tasks
were predominantly observed during the performance dip after 800s of runtime
discussed above.

\begin{figure}
  \up
  \centering
  \subfloat[][]{
    \includegraphics[width=0.44\textwidth]{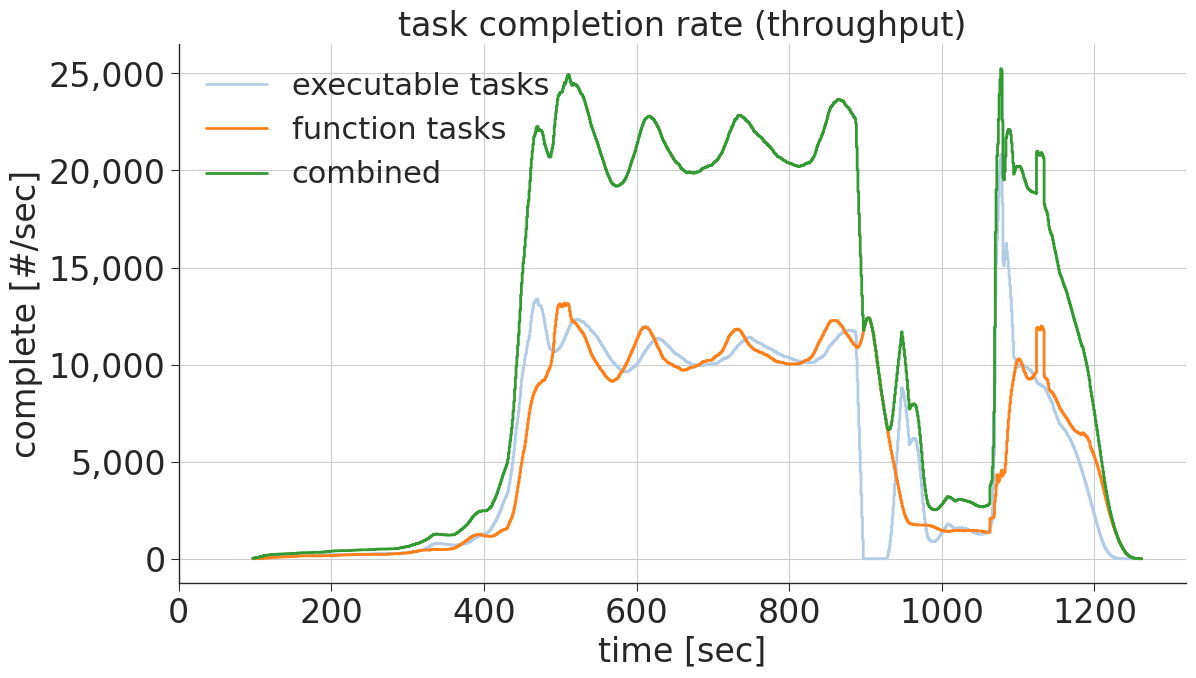}
    \label{sfig:exp3_rate}}
  \\
  \UP
  \subfloat[][]{
    \includegraphics[width=0.44\textwidth]{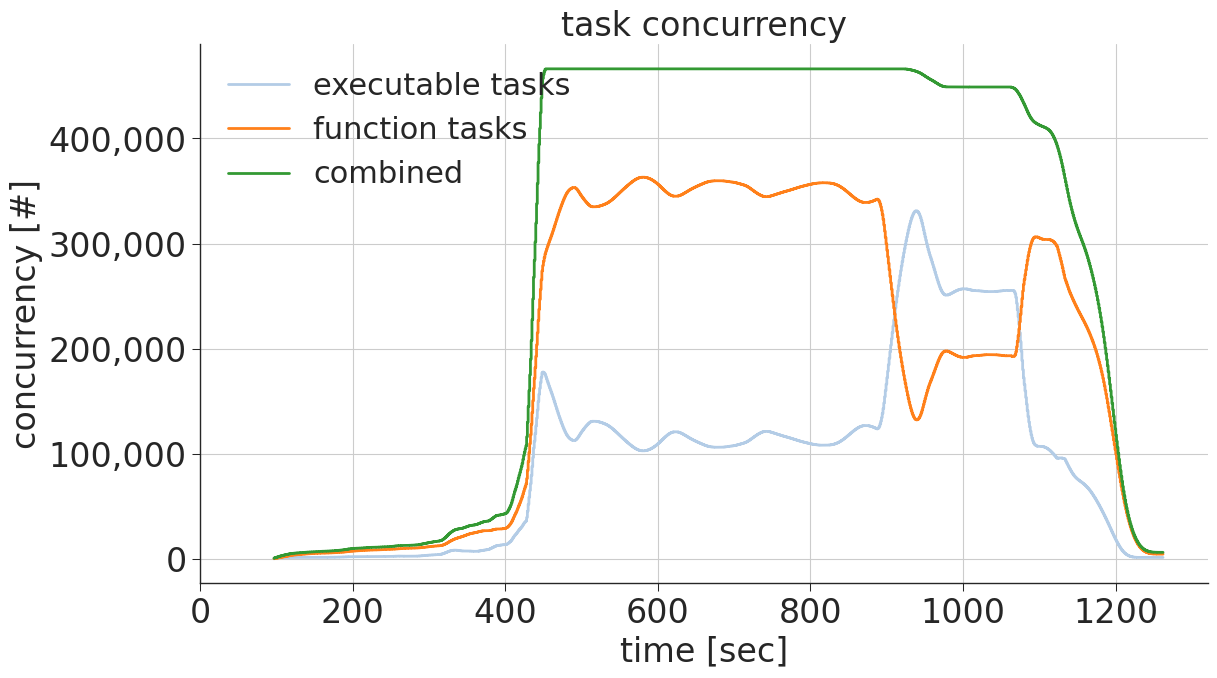}
    \label{sfig:exp3_concurrency}}
  \caption{Experiment 3: task completion rate and task concurrency.}
  \label{fig:exp3_concurrency}
  \UP
  \UP
\end{figure}

Fig.~\ref{sfig:exp3_rate} shows the task completion rate for experiment 3. The
rate shows a ramp up of $\sim$360s (see discussion above) followed by a peak of
$\sim$25,000 tasks/s and an average of 22,000 tasks/s for most of the remaining
time of the run. This is equivalent to a peak $\sim90\times10^6$ tasks/h and an
average $\sim79\times10^6$ tasks/h. The plot also shows the individual
completion rate of the executable and function tasks. Both rate and behavior
over time are comparable for function and executable tasks. After the ramp up
phase, the execution rate peaks at $\sim$13,000 task/s to then stabilize at an
average of $\sim$11,000 task/s.

As discussed, we observe some stalling at around 800s and then a rate peak of
$\sim$25,000 task/s at $\sim$1000s that includes the tasks completed after the
stalling period. We observe an cooldown phase in which the number of remaining
tasks progressively decreases until no tasks are left to be executed. The
consistency of behavior for function and executable tasks indicates that RAPTOR
can concurrently execute both types of task in isolation, without affecting
overall performance.

\subsection{Experiment 4}\label{ssec:exp4}

\begin{figure}
  \up
  \centering
  \subfloat[][]{
    \includegraphics[width=0.44\textwidth]{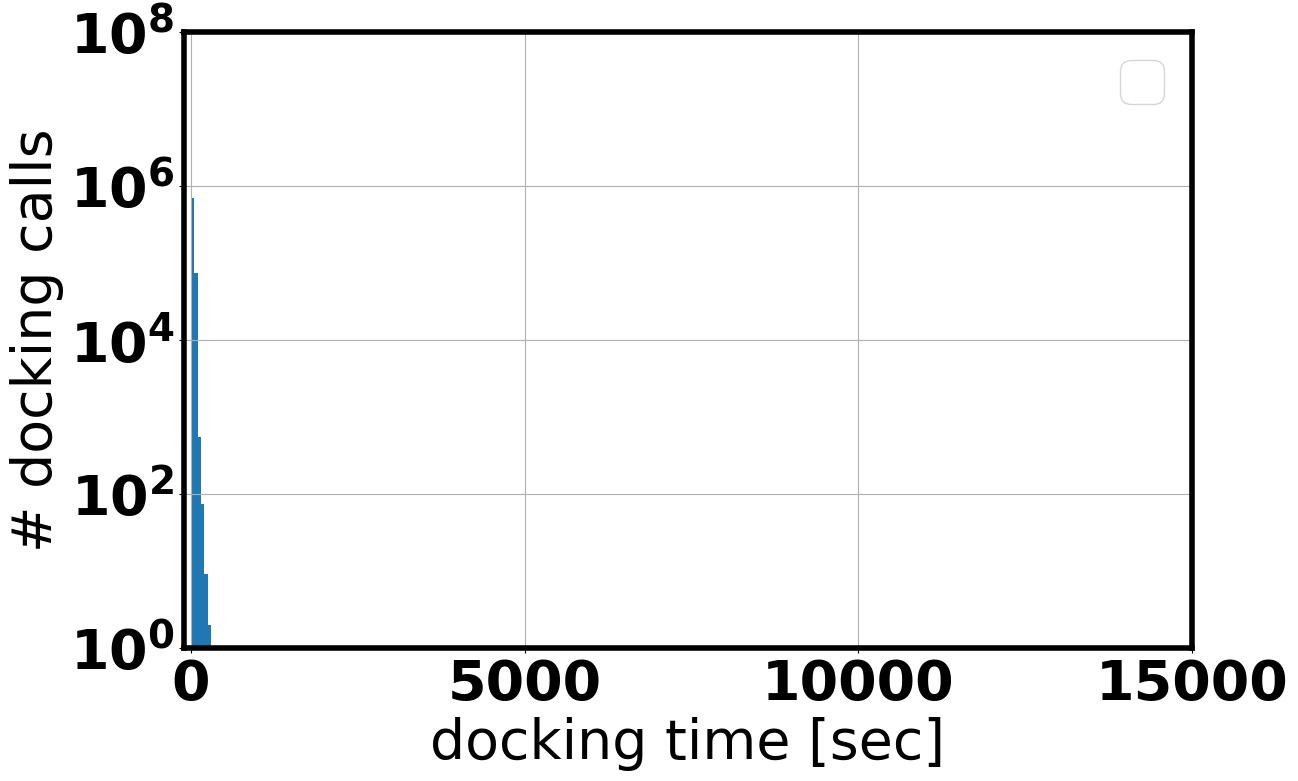}
    \label{sfig:exp4_time_summit_hero}}
  \\
  \UP
  \subfloat[][]{
    \includegraphics[width=0.44\textwidth]{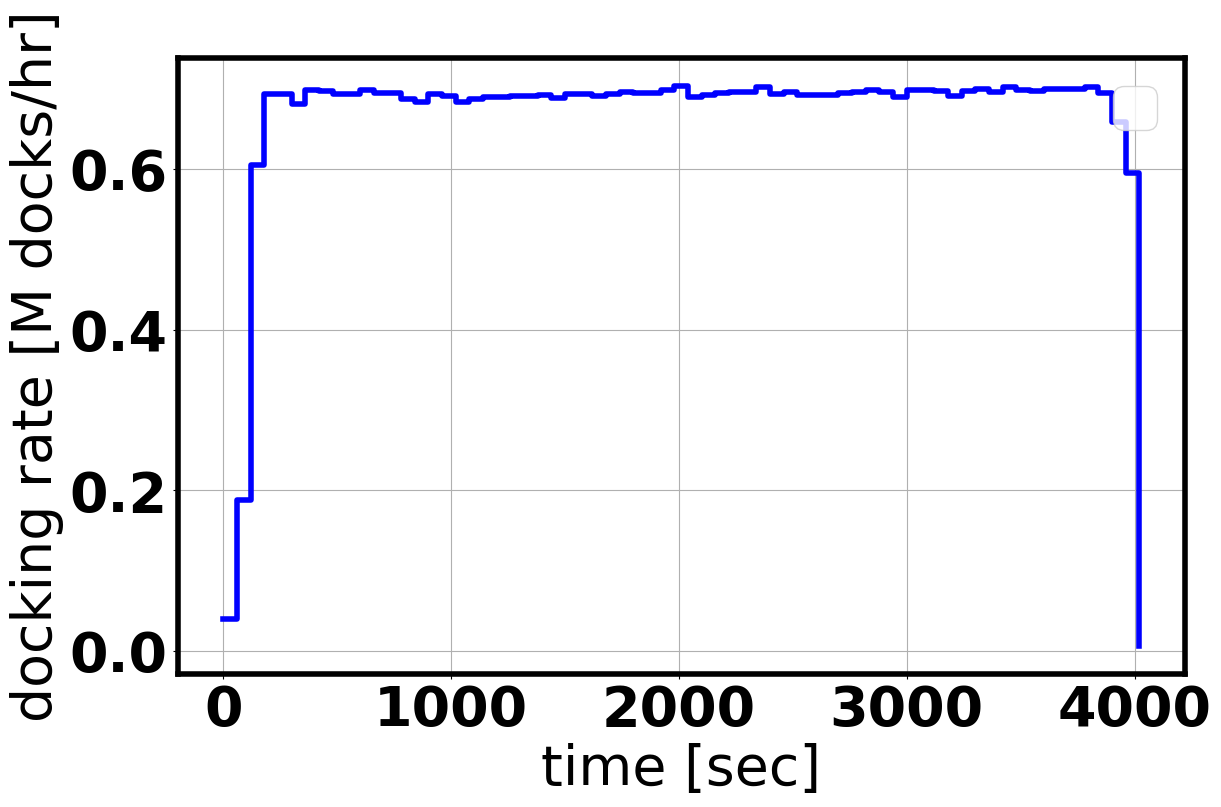}
    \label{sfig:exp4_rate_summit_hero}}
  \caption{Experiment 3: (a) Distribution of docking time and (b) docking rate
           for a single protein and $57\times10^6$ ligands. A pilot is
           concurrently executed on Summit with 6000 GPUs.}
  \label{fig:exp4_summit_hero}
  \UP
  \UP
\end{figure}

Figure~\ref{sfig:exp4_time_summit_hero} shows the distribution of docking times
of $\sim57\times10^6$ ligands from the
\texttt{mcule\--ultimate\--200\--204\--VJL} database, using AutoDock-GPU on
Summit. The distribution has a max/mean of 263.9/36.2s (Tab.~\ref{tab:exp}).
Compared to experiment 1, Fig.~\ref{fig:exp1_durations}, max docking time is
shorter, but the mean is longer. Compared to experiments 2 and 3, both max and
mean are shorter. As observed, those differences are due to specific properties
of the docked ligands and the protein.

Fig.~\ref{sfig:exp4_rate_summit_hero} shows the docking rate for one pilot with
1000 nodes (6000 GPUs). Different from experiments 1, 2 and especially 3, the
rate peaks very rapidly at $\sim11\times10^6$ docks/h, showing a very short
startup time. RAPTOR maintains that steady rate until the end of the execution,
with a very rapid cooldown phase compared to the other experiments. We explain
the observed sustained dock rate with an interplay between the scoring function
and its implementation in AutoDock-GPU, and specific features of the
$57\times10^6$ docked ligands.

Different from OpenEye on Frontera, AutoDock-GPU bundles 16 ligands into one GPU
computation in order to efficiently use the GPU memory, reaching an average
docking rate of $11.1\times10^6$ docks/h (Tab.~\ref{tab:exp}). Currently, our
profiling capabilities allow us to measure GPU utilization with 5\% relative
error. Based on our profiling, we utilized between 93 and 98\% of the available
GPU resources.

Experiment 4 shows how RAPTOR can manage resource and executable tasks
heterogeneities. Moving from Frontera to Summit required few changes in the
pilot description and a new task description for AutoDock-GPU \I{executable}
tasks instead of OpenEye \I{function} tasks. RAPTOR coordinator and worker also
required minimal configuration changes to account for different number of
resources available per node but, importantly, no change was required to manage
executable tasks instead of function tasks. Both RP and RAPTOR are agnostic
towards the type of code executed by each task.

Together, experiments 1--4 show how RAPTOR manages 6 types of heterogeneity: (1)
different number of pilots per run; (2) different types of tasks concurrently
executed on the same pilot, coordinator and worker; (3) different task runtimes;
(4) different type of task executable; (5) different type of HPC platform; and
(6) different type of computing resource, CPU cores and GPUs. Across these
heterogeneities, RAPTOR manages to reach more than 90\% peak resource
utilization and unprecedented scales for the docking calculations.

By supporting heterogeneity, flexible API and scale, RP and RAPTOR enable users
to write more general-purpose and flexible applications, abstracting resource
provisioning and management, and task scheduling. Ultimately, the goal is to
avoid having to write special purpose applications that support a single use
case on a specific platform and at a specific scale. This is what the COVID19
use case required: rapid deployment at scale on diverse platforms to efficiently
and effectively leverage all the computational capacity across multiple
institutions, to obtain results in the shortest possible amount of time.

%% file: related.tex


We discussed the performance and scale that RAPTOR achieves for the
computational docking problem relative to similar efforts
(Sec.~\S\ref{sec:docking}) that represent the state-of-the-art. RAPTOR
achieves at least a factor of two greater throughput on any platform than
published results. We attribute this to the combination of the of the
coordinator/worker and pilot paradigms~\cite{turilli2018comprehensive} and their
scalable implementations.

Coordinator/worker is one of the most common paradigms in distributed computing and
programming~\cite{goux2000metacomputing}. Many classes of algorithms naturally
fit the coordinator/worker paradigm, making it useful both for writing user-facing
applications and scheduler components for middleware systems, especially in
presence of heterogeneous resources and a dynamic runtime environment.
Coordinator/worker is commonly adopted by middleware to enable the execution of
many-task applications on distributed computing
infrastructures~\cite{goux2000enabling}.

In HPC, coordinator/worker is commonly used to coordinate the concurrent execution of
processes and tasks across multiple resources and compute nodes. At programming
level, coordinator/worker is used with MPI
libraries~\cite{rynge2012enabling,reussner1998skampi} and language extensions
like Charm++~\cite{kale1993charm++,langer2012performance} or
COMPSs~\cite{conejero2018task}, to implement large-scale, single-executable
applications. At task-level, diverse frameworks use workers coordinated by a
coordinator to distribute and then execute tasks across HPC resources. For example,
Dask~\cite{rocklin2015dask}, Parsl~\cite{babuji2019parsl},
Spark~\cite{zaharia2010spark} and Arkouda~\cite{merrill2019arkouda} all use the
coordinator/worker paradigm but many single-point solutions for domain-specific use
cases also use
coordinator/worker~\cite{guo2015fault,young2017evolving,laguna2016evaluating}.

While in this paper we present experiments specific to the problem described in
Sec.~\S\ref{sec:docking}, via the , RAPTOR supports the development of
domain-independent applications with homogeneous and heterogeneous tasks. This
is because the coordinator/worker paradigm is domain-independent and RAPTOR poses no
constraints on the type of computation performed by the functions or executable
of the workload. Further, RAPTOR supports extreme scale on HPC platforms with
diverse architectures and resource usage policies.

%% file: conclusions.tex


HTVS pipelines are used in a variety of disciplines, ranging from materials
design~\cite{pyzer2015high} to molecular design~\cite{zhang2008dovis}. In
particular, multi-scale biophysics-based HTVS pipelines are an important
strategy for computational drug development. While HTVS can be considerably
faster than experimental screening, until now, it has been too slow to explore
libraries with billions of molecules, even on the fastest machines.

In this paper we describe the design and implementation of RAPTOR, and offer a
performance evaluation of RAPTOR when used to perform computational docking at
scale. Docking is often the first stage of multi-scale, biophysics-based HTVS
pipelines and we report progress towards addressing the performance challenges
of computational docking at scale. This, in turn, provides a path towards an
overall improvement of throughput of computational drug discovery, in terms of
size of libraries screened, and the possibility of integrating machine learning
components with physics-based components.

RAPTOR extends the Pilot abstraction with the coordinator/worker paradigm, and
offers a general-purpose, task-level application programming interface (API) to
code distributed applications. RAPTOR supports: executing function and
executable tasks; achieving high throughput and high resource utilization with
arbitrary short running tasks; arbitrary partitioning of resources and tasks;
multilevel scheduling in which workloads are partitioned and then subsets of
tasks are locally scheduled to subset of resources; and partitioning of tasks
across multiple, independent executors.

RAPTOR offers three main advantages compared to existing frameworks for
multi-task applications: (1) users do not have to explicitly manage task
concurrency when coding multi-task workloads; (2) user applications can execute
up to $100\times10^6$ arbitrary Python functions on up to $500\times10^3$ cores
and $24\times10^3$ GPUs; and (3) users can avoid coding resource management and
task coordination. Additionally, RAPTOR abstracts away the notion of task
concurrency from the users, while supporting unprecedented scale and managing 6
types of heterogeneity: (1) number of pilots per run; (2) types of tasks
concurrently executed on the same pilot, coordinator and worker; (3) task
runtimes; (4) type of task executable; (5) type of HPC platform; and (6) type of
computing resource, CPU cores and GPUs. Overall, RAPTOR enables pilot-based
execution of multi-task workloads on most DoE and NSF HPC platforms, independent
of the type of executable launched by each task and, ultimately, of the use case
supported.

RAPTOR allows users to implement rich control logic for their applications,
expressed as an implementation of the coordinator/worker pattern. This will
become increasingly important as emerging platforms offer progressively more
heterogeneous resources and execution environments. Consistently, we plan to
extend RAPTOR with workload management features such as: enacting failure
management policies; making decisions based on state or output of tasks, both
during runtime or after task completion; satisfy data dependencies, enabling
both data and control flow management.